\documentclass[10pt,twocolumn,english]{revtex4}
\usepackage[T1]{fontenc}
\usepackage[latin9]{inputenc}
\usepackage{color}
\usepackage{array}
\usepackage{mathtools}
\usepackage{multirow}
\usepackage{amsmath}
\usepackage{amsthm}
\usepackage{amssymb}
\usepackage{graphicx}

\makeatletter

\providecommand{\tabularnewline}{\\}

\@ifundefined{textcolor}{}
{%
 \definecolor{BLACK}{gray}{0}
 \definecolor{WHITE}{gray}{1}
 \definecolor{RED}{rgb}{1,0,0}
 \definecolor{GREEN}{rgb}{0,1,0}
 \definecolor{BLUE}{rgb}{0,0,1}
 \definecolor{CYAN}{cmyk}{1,0,0,0}
 \definecolor{MAGENTA}{cmyk}{0,1,0,0}
 \definecolor{YELLOW}{cmyk}{0,0,1,0}
}

\newcommand{\Ket}[1]{{%
  \footnotesize\arraycolsep=0.3\arraycolsep\ensuremath{\left|\begin{matrix}#1\end{matrix}\right\rangle}}}

\newcommand{\updwn}{\rotatebox[origin=c]{90}{\footnotesize $\leftrightharpoons$}}
\newcommand{\supdwn}{\rotatebox[origin=c]{90}{\scalebox{.5}{$\leftrightharpoons$}}}

\makeatother

\usepackage{babel}
\begin{document}
\title{Low temperature pseudo-phase transition in an extended Hubbard diamond
chain}
\author{Onofre Rojas}
\affiliation{Departamento de Física, Universidade Federal de Lavras, 37200-900,
Lavras - MG, Brazil}
\author{Jordana Torrico,}
\affiliation{Departamento de Física, Universidade Federal de Minas Gerais, C. P.
702, 30123-970, Belo Horizonte - MG, Brazil.}
\author{L. M. Veríssimo}
\affiliation{Instituto de Física, Universidade Federal de Alagoas, 57072-970 Maceió
- AL, Brazil}
\author{M. S. S. Pereira}
\affiliation{Instituto de Física, Universidade Federal de Alagoas, 57072-970 Maceió
- AL, Brazil}
\author{S. M. de Souza}
\affiliation{Departamento de Física, Universidade Federal de Lavras, 37200-900,
Lavras - MG, Brazil}
\author{M. L. Lyra}
\affiliation{Instituto de Física, Universidade Federal de Alagoas, 57072-970 Maceió
- AL, Brazil}
\begin{abstract}
We consider the extended Hubbard diamond chain with an arbitrary number
of particles driven by chemical potential. The interaction between
dimer diamond chain and nodal couplings is considered in the atomic limit (no
hopping), while the dimer interaction includes the hopping term. We demonstrate
that this model exhibits a pseudo-transition effect in the low-temperature
regime. Here, we explore the pseudo-transition rigorously by analyzing
several physical quantities. The internal energy and entropy depict
sudden, although continuous, jumps which closely resembles discontinuous
or first-order phase transition. At the same time, the correlation
length and specific heat exhibit astonishing strong sharp peaks, quite
similar to a second-order phase transition. We
associate the ascending and descending part of the peak with  power-law
\textquotedbl pseudo-critical\textquotedbl exponents. We determine the pseudo-critical
exponents in the temperature range where these peaks are developed,
namely $\nu=1$ for the correlation length and $\alpha=3$ for the
specific heat. We also study the behavior of the electron density
and isothermal compressibility around the pseudo-critical temperature. 
\end{abstract}
\maketitle

\section{\label{sec:intro}Introduction}

In recent investigations of several decorated one-dimensional models with short-range interactions,
the first derivative of free energy, such as entropy, internal energy,
and magnetization, shows a steep like function of temperature but
still with continuous change which is quite similar to a first-order
phase transition behavior. The second-order derivative of free energy,
like the specific heat and magnetic susceptibility, resembles a typical
second-order phase transition behavior at finite temperature. This
peculiar behavior drew attention to a more careful study, as considered
in reference \citep{pseudo}. In reference\citep{Isaac}, an additional
discussion of the above phenomenology focused in the behavior of correlation
function for arbitrarily distant spins around the pseudo-transition.
Similar pseudo-transitions were shown to take place in the Ising-Heisenberg
diamond chain\citep{torrico,torrico2} and even in the pure Ising
diamond chain\citep{Strecka-ising}. It also has been explored in
the one-dimensional double-tetrahedral model, where the nodal sites
are occupied by localized Ising spins and alternate with a pair of
delocalized mobile electrons within a triangular plaquette\citep{Galisova}.
 Similarly, ladder model with alternating Ising-Heisenberg coupling\citep{on-strk},
as well as the triangular tube model with Ising-Heisenberg coupling
\citep{strk-cav} depict pseudo-transition signatures. A universal character of pre-asymptotic pseudo-critical exponents has been demonstrated\cite{unv-cr-exp,hutak21}. These pseudo-transitions taking place at finite-temperatures in one-dimensional model systems with short-range interactions are of a distinct nature from the true phase-transition exhibited in the presence of long-range couplings for which the correlation lenght diverges, but e.g
the specific heat can be without divergence \citep{daruka}.
In all the
above model systems presenting a pseudo-transition at finite temperature,
one of the couplings was assumed to be Ising-like in order to allow
for the exact calculation of the thermodynamic quantities. Examples
of pseudo-transitions taking place in one-dimensional systems of interacting
electrons without the assumption of an Ising-like nature of relevant
couplings are still missing.

The Hubbard model is one of the simplest models that describe more
accurately strongly interacting electron systems, which have attracted
a great deal of interest over the past decades related to the possible
emergence of geometrical frustration properties \citep{hagemann,lierop}.
Magnetic frustration in highly correlated electron models arises due
to the geometric structure of the lattice, which induces system failures
to satisfy simultaneously conflicting local requirements. The geometric
frustration of Hubbard model has been extensively studied, particularly
in the diamond chain structure, as considered by Derzhko et al.\citep{Derzhko09,Derzhko10,oleg14},
where frustration for a particular class of lattice was discussed.
Montenegro-Filho and Coutinho-Filho\citep{montenegro06} also considered
the doped $AB_{2}$ Hubbard chain, both in the weak coupling and the
infinite-$U$ limit (atomic limit) where quite interesting phases
were identified as a function of hole doping away from the half-filled
band, as well as 1/3-plateau magnetization, Kosterlitz-Thouless transition,
and Luttinger liquid\citep{montenegro20}. Further, Gulacsi et al.\citep{gulacsi}
also discussed the diamond Hubbard chain in a magnetic field and a
wide range of properties such as flat-band ferromagnetism, correlation-induced
metallic and half-metallic processes. The thermodynamics of the Hubbard
model on a diamond chain in the atomic limit was discussed in reference
\citep{hubd-dmd-ch}. Furthermore, frustrated quantum Heisenberg double-tetrahedral
and octahedral chains at high magnetic fields was discussed in\citep{krupnitska}.
Fermionic entanglement due to spin frustration was investigated in
hybrid diamond chain with localized Ising spins and mobile electrons\citep{torrico16}.

On the other hand, generally rigorous analysis of the Hubbard model
is a challenging task. Only in a particular case it is possible to
obtain exact results \citep{elliot}. Earlier in the seventies, Beni
and Pincus\citep{beni-pincus} focused in the one-dimensional Hubbard
model. Later Mancini \citep{Mancini,mancini2008} discussed several
additional properties of extended one-dimensional Hubbard model in
the atomic limit, obtaining the chemical potentials plateaus of the
particle density, as a function of the on-site Coulomb potential at
zero temperature. Earlier, the spinless versions of the Hubbard model
on diamond chain also was investigated\citep{spinless}, as well as
Lopes and Dias\citep{Lopes} performed a detailed investigation using
the exact diagonalization approach. The Ising-Hubbard diamond chain
has been investigated in reference \citep{lisnii}. In addition, experimental
data regarding the $1/3$ magnetization plateau in azurite\citep{rule,kikuchi}
were reproduced in several theoretical model systems such as the Ising-Heisenberg
diamond chain\citep{canova06,ananikian,chakh}. The quantum block-block
entanglement was investigated in the one-dimensional extended Hubbard
model by exact diagonalization\citep{sh-deng}. When the absolute
value of the nearest-neighbor Coulomb interaction becomes small, the
effects of the hopping term and the on-site interaction cannot be
neglected. The experimental observation of the double peaks both in
the magnetic susceptibility and specific heat \citep{Pereira08,jeschke11,Bo-Gu}
can be described accurately by the extended Hubbard diamond chain
model without the hopping of electrons or holes between the nodal
sites.

From an experimental point of view, the diamond chain
structure is also motivating. Recently, the compound $\mathrm{Cu_{3}(CH_{3}COO)_{4}(OH)_{5}\cdot5H_{2}O}$
has been synthetized \citep{cui19}, which exhibits a unique one-dimensional
diamond chain structure. There are other compounds such as $\mathrm{Cu_{3}Cl_{6}(H_{2}O)_{2}\cdot2H_{8}C_{4}SO_{2}}$
trimer chain system\citep{Honecker01}, as well as the well known natural mineral azurite
$\mathrm{Cu_{3}(CO_{3})_{2}(OH)_{2}}$ \citep{jeschke11} which are well represented by 1D diamond
chains. These and similar compounds would be ideal physical systems on which pseudo-transitions at finite temperature could be
searched.

Here, we advance in the study of pseudo-phase transitions in quasi
one-dimensional systems by presenting a detailed exact study of the
thermodynamic properties of the diamond chain Hubbard model in the
atomic limit. The present article is organized as follow: In sec.2,
we revisit the extended Hubbard model on diamond chain structure\citep{hubd-dmd-ch}
with nodal sites considered in the atomic limit. In sec.3 we present
our main findings where we focus in the existence of a pseudo-critical
temperature by exploring the behavior of the correlation length. In
sec.4 (sec.5) we analyze first (second) derivative physical quantities
in the vicinity of the pseudo-transition. Finally, sec. 6 is devoted
to our conclusions and perspectives.

\section{\label{sec:Extended}The Extended Hubbard model}

In this section, we revisit the model considered in reference \citep{hubd-dmd-ch}.
Some results that will be used in the following section are updated
and summarized. The model illustrated in fig.\ref{fig:extd-hubbrd-diamd},
consider the hopping term $t$ between sites $a$ and $b$. Additionally,
there is an onsite Coulomb repulsion interaction $U$ and nearest
neighbor repulsion interaction $V$ between $a$ and $b$, whereas
$V_{1}$ corresponds the coupling between of nodal sites $c$ with
sites $a$ and $b$ (as
labeled in the last block of fig.\ref{fig:extd-hubbrd-diamd}). We also assume that this model has an arbitrary
particle density. Thus, the system will be described by including
a chemical potential denoted by $\mu$. The Hamiltonian of the proposed
model can be expressed by:

\begin{figure}
\includegraphics[scale=0.8]{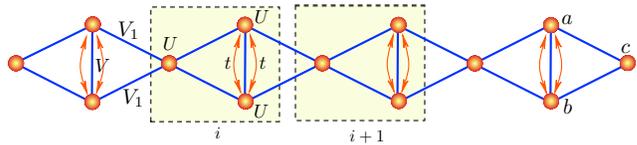}
\caption{\label{fig:extd-hubbrd-diamd}Schematic representation of the extended
Hubbard model on the diamond chain. Onsite Coulomb repulsion interaction
is denoted by $U$ and nearest neighbor repulsion interaction is represented
by $V$ and $V_{1}$, $t$ stands for the electron hopping term. }
\end{figure}

\begin{equation}
\boldsymbol{H}=\sum_{i=1}^{N}\boldsymbol{H}_{i,i+1},
\end{equation}
with $N$ being the number of unit cells (sites $a$, $b$ and $c$),
and $\boldsymbol{H}_{i,i+1}$ is given by

\begin{alignat}{1}
\boldsymbol{H}_{i,i+1}= & -t\sum\limits _{\sigma=\downarrow,\uparrow}{\normalcolor \left(\boldsymbol{a}_{i,\sigma}^{\dagger}\boldsymbol{b}_{i,\sigma}+{\color{red}{\normalcolor \boldsymbol{b}_{i,\sigma}^{\dagger}\boldsymbol{a}_{i,\sigma}}}\right)}\nonumber \\
 & -\mu\left(\boldsymbol{n}_{i}^{a}+\boldsymbol{n}_{i}^{b}+\boldsymbol{n}_{i}^{c}\right)\nonumber \\
 & +U\left(\boldsymbol{n}_{i,\uparrow}^{a}\boldsymbol{n}_{i,\downarrow}^{a}+\boldsymbol{n}_{i,\uparrow}^{b}\boldsymbol{n}_{i,\downarrow}^{b}+\boldsymbol{n}_{i,\uparrow}^{c}\boldsymbol{n}_{i,\downarrow}^{c}\right)\nonumber \\
 & V\boldsymbol{n}_{i}^{a}\boldsymbol{n}_{i}^{b}+V_{1}(\boldsymbol{n}_{i}^{a}+\boldsymbol{n}_{i}^{b})(\boldsymbol{n}_{i}^{c}+\boldsymbol{n}_{i+1}^{c}),\label{eq:Hamt-dmnd}
\end{alignat}
with $\boldsymbol{a}_{i,\sigma}$, and $\boldsymbol{b}_{i,\sigma}$
($\boldsymbol{a}_{i,\sigma}^{\dagger}$ and $\boldsymbol{b}_{i,\sigma}^{\dagger}$)
being the Fermi annihilation (creation) operators for electrons, while
$\sigma$ stands for the electron spin, and $\boldsymbol{n}_{i,\sigma}^{\alpha}$
stands for the number operator, with $\alpha=\{a,b,c\}$. Using
this number operator, we also define conveniently the following operators
$\boldsymbol{n}_{i}^{\alpha}=\boldsymbol{n}_{i,\uparrow}^{\alpha}+\boldsymbol{n}_{i,\downarrow}^{\alpha}$.

In order to contract and symmetrize the Hamiltonian (\ref{eq:Hamt-dmnd}),
we can define properly the following operators 
\begin{equation}
\begin{array}{rcl}
\boldsymbol{p}_{i,i+1} & = & \tfrac{1}{2}(\boldsymbol{n}_{i}^{c}+\boldsymbol{n}_{i+1}^{c}),\\
\boldsymbol{q}_{i,i+1} & = & \tfrac{1}{2}(\boldsymbol{n}_{i,\uparrow}^{c}\boldsymbol{n}_{i,\downarrow}^{c}+\boldsymbol{n}_{i+1,\uparrow}^{c}\boldsymbol{n}_{i+1,\downarrow}^{c}).
\end{array}
\end{equation}

Using these operators, we can rewrite the Hamiltonian (\ref{eq:Hamt-dmnd}),
which becomes as follows:

\begin{alignat}{1}
{\color{black}\boldsymbol{H}_{i,i+1}=} & {\color{black}-t\sum\limits _{\sigma=\downarrow,\uparrow}\left(\boldsymbol{a}_{i,\sigma}^{\dagger}\boldsymbol{b}_{i,\sigma}+{\color{red}{\normalcolor \boldsymbol{b}_{i,\sigma}^{\dagger}\boldsymbol{a}_{i,\sigma}}}\right)-\mu\boldsymbol{p}_{i,i+1}}\nonumber \\
 & -(\mu-2V_{1}\boldsymbol{p}_{i,i+1})\left(\boldsymbol{n}_{i}^{a}+\boldsymbol{n}_{i}^{b}\right)+V\boldsymbol{n}_{i}^{a}\boldsymbol{n}_{i}^{b}\nonumber \\
 & +U\left(\boldsymbol{n}_{i,\uparrow}^{a}\boldsymbol{n}_{i,\downarrow}^{a}+\boldsymbol{n}_{i,\uparrow}^{b}\boldsymbol{n}_{i,\downarrow}^{b}\right)+U\boldsymbol{q}_{i,i+1}.\label{eq:Ham-red}
\end{alignat}

It is worth mentioning that this model already was investigated in
reference \citep{hubd-dmd-ch}, for arbitrary number of electrons.
Here we consider in each site the following basis $\{0,\uparrow,\downarrow,\updwn\}$.

The eigenvalues and eigenvectors of the dimer plaquette are summarized
in Table \ref{tab: eig-vals}, which are valid
in general for arbitrary values of the Hamiltonian \eqref{fig:extd-hubbrd-diamd}
parameters. 

\begin{table*}
\begin{tabular}[t]{|c|c|l|c|l|}
\hline 
$m$ & $M_{ab}$ & Eigenvalues & $g$ & Eigenvectors\tabularnewline
\hline 
\hline 
\noalign{\vskip\doublerulesep}
$0$ & $0$ & $\lambda_{0,0}=\boldsymbol{D}_{0},$ & $1$ & $|S_{00}\rangle=\Ket{\begin{array}{c}
0\\
0
\end{array}}$\tabularnewline[2.5mm]
\hline 
\noalign{\vskip\doublerulesep}
$1$ & $0.5$ & $\lambda_{0,\sigma}^{(\pm)}=\boldsymbol{D}_{1}\pm t$ & $\begin{array}{c}
2\\
2
\end{array}$ & $|S_{0\sigma}^{(\pm)}\rangle=\tfrac{1}{\sqrt{2}}\left(\Ket{\begin{array}{c}
0\\
\sigma
\end{array}}\mp\Ket{\begin{array}{c}
\sigma\\
0
\end{array}}\right)$\tabularnewline[2.5mm]
\hline 
\noalign{\vskip\doublerulesep}
\multirow{5}{*}{$2$} & $1$ & $\lambda_{\sigma,\sigma}=\boldsymbol{D}_{2}+V$ & $2$ & $|S_{\sigma\sigma}\rangle=\Ket{\begin{array}{c}
\sigma\\
\sigma
\end{array}}$\tabularnewline[2.5mm]
\cline{2-5} \cline{3-5} \cline{4-5} \cline{5-5} 
\noalign{\vskip\doublerulesep}
 & $0$ & $\lambda_{{\scriptscriptstyle 0,\supdwn}}^{(1)}=\boldsymbol{D}_{2}+U$ & $1$ & $|S_{0,\supdwn}^{(1)}\rangle=\tfrac{1}{\sqrt{2}}\left(\Ket{\begin{array}{c}
\updwn\\
0
\end{array}}-\Ket{\begin{array}{c}
0\\
\updwn
\end{array}}\right)$\tabularnewline[2.5mm]
\cline{2-5} \cline{3-5} \cline{4-5} \cline{5-5} 
\noalign{\vskip\doublerulesep}
 & $0$ & $\lambda_{{\scriptscriptstyle \updownarrow\updownarrow}}^{(+)}=\boldsymbol{D}_{2}+V+2t\cot(\theta)$ & $1$ & $|S_{{\scriptscriptstyle \updownarrow\updownarrow}}^{(+)}\rangle=\tfrac{1}{\sqrt{2}}\left\{ \cos(\theta)\left(\Ket{\begin{array}{c}
\updwn\\
0
\end{array}}+\Ket{\begin{array}{c}
0\\
\updwn
\end{array}}\right)-\sin(\theta)\left(\Ket{\begin{array}{c}
\uparrow\\
\downarrow
\end{array}}+\Ket{\begin{array}{c}
\downarrow\\
\uparrow
\end{array}}\right)\right\} $\tabularnewline[2.5mm]
\cline{2-5} \cline{3-5} \cline{4-5} \cline{5-5} 
\noalign{\vskip\doublerulesep}
 & $0$ & $\lambda_{{\scriptscriptstyle \updownarrow\updownarrow}}^{(-)}=\boldsymbol{D}_{2}+V-2t\tan(\theta)$ & $1$ & $|S_{{\scriptscriptstyle \updownarrow\updownarrow}}^{(-)}\rangle=\tfrac{1}{\sqrt{2}}\left\{ \sin(\theta)\left(\Ket{\begin{array}{c}
\updwn\\
0
\end{array}}+\Ket{\begin{array}{c}
0\\
\updwn
\end{array}}\right)+\cos(\theta)\left(\Ket{\begin{array}{c}
\uparrow\\
\downarrow
\end{array}}+\Ket{\begin{array}{c}
\downarrow\\
\uparrow
\end{array}}\right)\right\} $\tabularnewline[2.5mm]
\cline{2-5} \cline{3-5} \cline{4-5} \cline{5-5} 
\noalign{\vskip\doublerulesep}
 & $0$ & $\lambda_{{\scriptscriptstyle \downarrow,\uparrow}}^{(2)}=\boldsymbol{D}_{2}+V$ & $1$ & $|S_{{\scriptscriptstyle \downarrow,\uparrow}}^{(2)}\rangle=\tfrac{1}{\sqrt{2}}\left(\Ket{\begin{array}{c}
\uparrow\\
\downarrow
\end{array}}-\Ket{\begin{array}{c}
\downarrow\\
\uparrow
\end{array}}\right)$\tabularnewline[2.5mm]
\hline 
\noalign{\vskip\doublerulesep}
$3$ & $0.5$ & $\lambda_{\supdwn,\sigma}^{(\pm)}=\boldsymbol{D}_{3}+U+2V\pm t$ & $\begin{array}{c}
2\\
2
\end{array}$ & $|S_{\supdwn,\sigma}^{(\pm)}\rangle=\tfrac{1}{\sqrt{2}}\left(\Ket{\begin{array}{c}
\updwn\\
\sigma
\end{array}}\mp\Ket{\begin{array}{c}
\sigma\\
\updwn
\end{array}}\right)$\tabularnewline[2.5mm]
\hline 
\noalign{\vskip\doublerulesep}
$4$ & $0$ & $\lambda_{{\scriptscriptstyle \supdwn,\supdwn}}=\boldsymbol{D}_{4}+2U+4V$ & $1$ & $|S_{{\scriptscriptstyle \supdwn,\supdwn}}\rangle=\Ket{\begin{array}{c}
\updwn\\
\updwn
\end{array}}$\tabularnewline[2.5mm]
\hline 
\end{tabular}\caption{\label{tab: eig-vals}The first column means the number of particles
in dimer plaquette, the second column describes the dimer plaquette
magnetization, third column reports the eigenvalues, while fourth
column corresponds to the eigenvalues degeneracy. The fifth column
corresponds to the eigenvectors of dimer plaquette. Here $\boldsymbol{D}_{m}=-(\mu-2V_{1}\boldsymbol{p}_{i,i+1})m-\mu\boldsymbol{p}_{i,i+1}+U\boldsymbol{q}_{i,i+1}$
and $\cot\left(2\theta\right)=\frac{U-V}{4t}$.}
\end{table*}

It is worth to mention that all analysis performed
throughout this work will be done in the thermodynamic limit.
Finite size effects can be evaluated in a similar
way to that put forward in reference \citep{finite-chain}.

\subsection{Phase Diagram}

In order to analyze some relevant features, we will
focus in the more interesting case when the particle-hole symmetry
is satisfied which follows the restriction $V_{1}=V/2$. Under
this condition, for instance, we can analyze the half-filled band case,
which occurs under the following restriction for the chemical potential
$\mu=U/2+2V$. 

It is worthy to mention that, the Hamiltonian (\ref{eq:Hamt-dmnd}),
has 64 eigenvalues per diamond plaquette. The zero temperature phase
diagram analysis was already discussed in reference \citep{hubd-dmd-ch}.
Below we just give some ground-state energies relevant to the following
analysis of the pseudo transition features.

\begin{figure}
\includegraphics[scale=0.5]{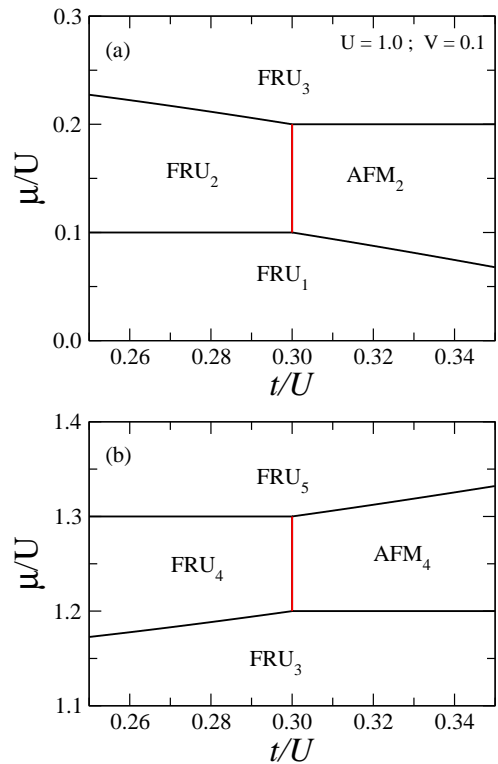}
\caption{\label{fig:Phase-diagram}Zero temperature phase diagram in the plane
$t-\mu$, assuming fixed $U=1$, $V=0.1$. (a) Shows the region of
$\mu/U$ values where $\rho=2/3$ phases appear; (b) The range of
$\mu/U$ values where $\rho=4/3$ phases appear.}
\end{figure}

Fig.\ref{fig:Phase-diagram}a illustrates the zero temperature phase
diagram in the plane $t-\mu$, for fixed $U=1$, $V=0.1$ and $0<\mu/U<0.3$.
There is a phase corresponding to a dimer antiferromagnetic ($AFM_{2}$)
state,

\begin{equation}
|AFM_{2}\rangle=\prod_{i=1}^{N}|S_{{\scriptscriptstyle \updownarrow\updownarrow}}^{(-)}\rangle 
|0\rangle.
\end{equation}
In this case, there are two electrons with opposite spins in the
dimer sites, while nodal sites are empty. Thus, the electron density
per unit cell is $\rho=2/3$. The corresponding eigenvalue is given
by 
\begin{equation}
E_{AFM_{2}}=-2\mu+V-2t\tan(\theta).
\end{equation}

There is also a phase corresponding to a dimer and nodal frustrated
state ($FRU_{2}$),

\begin{equation}
|FRU_{2}\rangle=\prod_{i=1}^{N}\tfrac{1}{\sqrt{2}}\left(\Ket{\begin{array}{c}
0\\
\sigma_{i}
\end{array}}-\Ket{\begin{array}{c}
\sigma_{i}\\
0
\end{array}}\right)|\tau_{i}\rangle.
\end{equation}
This state also has two electrons: one electron is in a dimer site
while the other occupies a nodal site, both with arbitrary spin orientation.
The electron density $\rho=2/3$. The corresponding residual entropy
in units of $k_{B}$\textcolor{red}{{} }is $\mathcal{S}=\ln(4)$, whereas
the ground-state energy becomes

\begin{equation}
E_{FRU_{2}}=-2\mu-t+V.
\end{equation}

The vertical red line corresponds to $t_c=(U-V)/3=0.3$. We will focus in this
phase boundary between $|AFM_{2}\rangle$ and $|FRU_{2}\rangle$.
The corresponding interface residual entropy is $\mathcal{S}=\ln(4)$,
as we will verify ahead.

Other surrounding states in the phase diagram are:

\begin{equation}
|FRU_{1}\rangle=\prod_{i=1}^{N}\tfrac{1}{\sqrt{2}}\left(\Ket{\begin{array}{c}
\sigma\\
0
\end{array}}-\Ket{\begin{array}{c}
0\\
\sigma
\end{array}}\right)|0\rangle\label{eq:fru1}
\end{equation}
with $n=1$ particle per unit cell and electron density $\rho=1/3$.

\begin{equation}
|FRU_{3}\rangle=\prod_{i=1}^{N}\tfrac{1}{\sqrt{2}}\left(\Ket{\begin{array}{c}
\updwn\\
0
\end{array}}-\Ket{\begin{array}{c}
0\\
\updwn
\end{array}}\right)|\tau_{i}\rangle\label{eq:fru3}
\end{equation}
with $n=3$ electrons per unit cell or $\rho=1$. The residual entropy
in the frustrated phases \eqref{eq:fru1} and \eqref{eq:fru3} is
given by $\mathcal{S}=\ln(2)$.

Similarly, in fig.\ref{fig:Phase-diagram}b we show the phase diagram
for the same set of fixed parameters but in the range $1.1<\mu/U<1.4$
where $\rho=4/3$ phases appear.

In this region we observe also another dimer antiferromagnetic ($AFM_{4}$)
state given by

\begin{equation}
|AFM_{4}\rangle=\prod_{i=1}^{N}|S_{{\scriptscriptstyle \updownarrow\updownarrow}}^{(-)}\rangle|\updwn\rangle,
\end{equation}
where two electrons with opposite spins are located in  the
dimer sites, and the other pair of electrons is located in the nodal
site. The respective ground-state energy becomes

\begin{equation}
E_{AFM_{4}}=-4\mu+5V+U-2t\tan(\theta).
\end{equation}

There is also a dimer and nodal frustrated state ($FRU_{4}$),

\begin{equation}
|FRU_{4}\rangle=\prod_{i=1}^{N}\tfrac{1}{\sqrt{2}}\left(\Ket{\begin{array}{c}
\updwn\\
\sigma_{i}
\end{array}}-\Ket{\begin{array}{c}
\sigma_{i}\\
\updwn
\end{array}}\right)|\tau_{i}\rangle,
\end{equation}
with corresponding ground-state energy 
\begin{equation}
E_{FRU_{4}}=-4\mu-t+5V+U,
\end{equation}
and electron density $\rho=4/3$.

The vertical red line corresponds to $t=(U-V)/3=0.3$. The phase boundary
between $|AFM_{2}\rangle$ and $|FRU_{2}\rangle$ on which the residual
entropy is $\mathcal{S}=\ln(4)$, as we will verify ahead.

The additional phase states illustrated in this diagram is composed
by $n=5$:

\begin{equation}
|FRU_{5}\rangle=\prod_{i=1}^{N}\tfrac{1}{\sqrt{2}}\left(\Ket{\begin{array}{c}
\updwn\\
\sigma_{i}
\end{array}}-\Ket{\begin{array}{c}
\sigma_{i}\\
\updwn
\end{array}}\right)|\updwn\rangle,
\end{equation}
with $\rho=5/3$. This frustrated phase has residual entropy $\mathcal{S}=\ln(2)$.

Further details of the ground state phase diagrams can be found in
reference\citep{hubd-dmd-ch}. Pseudo-transitions are identified in the close vicinity of the $AFM_{2}-FRU_{2}$ and $AFM_{4}-FRU_{4}$ phase-boundaries.

\subsection{\label{sec:Exact}Thermodynamics}

In order to use the decoration transformation approach, we write the
Boltzmann factors for the extended Hubbard model on diamond chain
as follows:

\begin{alignat}{1}
w_{\boldsymbol{n}_{i}^{c},\boldsymbol{n}_{i+1}^{c}}= & {\displaystyle \mathrm{e}^{-\beta\boldsymbol{D}_{0}}+4\left(\mathrm{e}^{-\beta\boldsymbol{D}_{1}}+\mathrm{e}^{-\beta(\boldsymbol{D}_{3}+U+2V)}\right)\cosh(\beta t)}\nonumber \\
 & +{\displaystyle \mathrm{e}^{-\beta\boldsymbol{D}_{2}}\left(\mathrm{e}^{-\beta U}+3\mathrm{e}^{-\beta V}\right)+\mathrm{e}^{-\beta(\boldsymbol{D}_{4}+2U+4V)}}\nonumber \\
 & +\mathrm{e}^{-\beta(\boldsymbol{D}_{2}+V)}\left(\mathrm{e}^{-\beta\vartheta_{+}t/2}+\mathrm{e}^{-\beta\vartheta_{-}t/2}\right),\label{eq:orig-W}
\end{alignat}
with

\begin{equation}
\boldsymbol{D}_{m}=-(\mu-V\boldsymbol{p}_{i,i+1})m-\mu\boldsymbol{p}_{i,i+1}+U\boldsymbol{q}_{i,i+1}.
\end{equation}
Here, $\vartheta_{\pm}$ is defined by

\begin{equation}
\vartheta_{\pm}=\frac{U-V\pm\sqrt{(U-V)^{2}+16t^{2}}}{t}.
\end{equation}

To solve the effective Hubbard model with up to four-body coupling,
we can use the transfer matrix method\citep{baxter-book}, similarly
to that used in reference \citep{beni-pincus,spinless}. Therefore
the symmetric Hamiltonian by exchanging $i\rightarrow i+1$ and $i+1\rightarrow i$,
leads to a symmetric transfer matrix which can be expressed by:

\begin{equation}
{\bf W}=\left[\begin{array}{cccc}
w_{0,0} & w_{0,1} & w_{0,1} & w_{0,2}\\
w_{0,1} & w_{1,1} & w_{1,1} & w_{1,2}\\
w_{0,1} & w_{1,1} & w_{1,1} & w_{1,2}\\
w_{0,2} & w_{1,2} & w_{1,2} & w_{2,2}
\end{array}\right],\label{eq:transf-Mtx}
\end{equation}
where the elements of $\mathbf{W}$ is given by Eq.~(\ref{eq:orig-W}).
Let us define a convenient notation, 
\begin{equation}
\begin{array}{ccl}
w_{0,0}(x) & = & {\displaystyle 1+2x\left(1+\frac{x^{2}}{z^{4}y^{2}}\right)\left(\frac{1}{\gamma^{2}}+\gamma^{2}\right)+}\\
 &  & {\displaystyle +x^{2}\left(\frac{3}{z^{2}}+\frac{1}{y^{2}}+\frac{1}{yz\varsigma}+\frac{\varsigma}{yz}\right)+\frac{x^{4}}{y^{4}z^{8}},}
\end{array}\label{eq:w00}
\end{equation}
with $x=\mathrm{e}^{\beta\mu}$, $y=\mathrm{e}^{\frac{1}{2}\beta U}$,
$z=\mathrm{e}^{\frac{1}{2}\beta V}$, $\gamma=\mathrm{e}^{\frac{1}{2}\beta t}$
and $\varsigma=\mathrm{e}^{\frac{1}{2}\beta\sqrt{(U-V)^{2}+16t^{2}}}$.
All other Boltzmann factors could be expressed in terms of $w_{0,0}(x)$
defined by Eq.~(\ref{eq:w00}) as follows: 
\begin{equation}
w_{n_{1},n_{2}}(x)=\frac{x^{(n_{1}+n_{2})/2}}{y^{\lfloor n_{1}/2\rfloor+\lfloor n_{2}/2\rfloor}}w_{0,0}\left(\frac{x}{z^{n_{1}+n_{2}}}\right),\label{eq:wnn}
\end{equation}
by $\lfloor\ldots\rfloor$ we mean the floor function for any real
number.

In order to carry out a reduced transfer matrix, we use the symmetry
of the system. The proposed Hamiltonian is invariant with respect
to electron spin orientation in nodal sites (site c). Therefore, the
reduced transfer matrix becomes

\begin{equation}
{\bf V}=\left[\begin{array}{ccc}
w_{0,0} & \sqrt{2}\,w_{0,1} & w_{0,2}\\
\sqrt{2}\,w_{0,1} & 2\,w_{1,1} & \sqrt{2}\,w_{1,2}\\
w_{0,2} & \sqrt{2}\,w_{1,2} & w_{2,2}
\end{array}\right].\label{eq:V}
\end{equation}

The determinant of the reduced transfer matrix becomes a cubic equation
of the form 
\begin{equation}
\mathrm{det}({\bf V}-\Lambda)=\left(\Lambda^{3}+a_{3}\Lambda^{2}+a_{2}\Lambda+a_{1}\right)=0,\label{eq:cub-eq}
\end{equation}
where the coefficients become

\begin{alignat}{1}
a_{1}= & 2w_{0,0}w_{1,2}^{2}+2w_{0,2}^{2}w_{1,1}+2w_{0,1}^{2}w_{2,2}\nonumber \\
 & -2w_{0,0}w_{1,1}w_{2,2}-4w_{0,2}w_{0,1}w_{1,2},\nonumber \\
a_{2}= & 2w_{0,0}w_{1,1}+2w_{1,1}w_{2,2}+w_{0,0}w_{2,2}\nonumber \\
 & -2w_{0,1}^{2}-w_{0,2}^{2}-2w_{1,2}^{2},\nonumber \\
a_{3}= & -w_{0,0}-2w_{1,1}-w_{2,2}.\label{eq:coefs}
\end{alignat}

Consequently, the roots of the algebraic cubic equation may be expressed
as follow 
\begin{equation}
\Lambda_{j}=2\sqrt{Q}\cos\left(\tfrac{\phi-2\pi j}{3}\right)-\frac{1}{3}a_{3},\quad j=0,1,2,\label{eq:sol-cub}
\end{equation}
with

\begin{alignat}{1}
\phi & =\arccos\left(\tfrac{R}{\sqrt{Q^{3}}}\right),\\
Q & ={\displaystyle \frac{a_{3}^{2}-3a_{2}}{9},}\label{eq:Q real}\\
R & ={\displaystyle \frac{9a_{2}a_{3}-27a_{1}-2a_{3}^{3}}{54}.}\label{eq: R real}
\end{alignat}
It is verified that $Q>0$ in appendix \ref{Real-roots-of}, which
implies that all three roots must be different and real. We also analyze
which eigenvalues must be the largest and the lowest one. So, it is enough to restrict $0<\phi<\pi$ in the cubic solution without
loosing its general solution, as discussed
in appendix \ref{Real-roots-of}. Other intervals just exchange the
cubic root solutions, as illustrated in table \ref{tab:Cubic-r}. In the interval $0<\phi<\pi$, the eigenvalues are
ordered as $\Lambda_{0}>\Lambda_{1}>\Lambda_{2}$.

\section{\label{sec:Therm}Pseudo-critical temperature}

In this section we will analyze the anomalous thermodynamic property
of the present extended diamond chain Hubbard model in the atomic
limit\citep{hubd-dmd-ch}. In order to study the thermodynamic properties,
we will use the exact free energy ${\displaystyle f=-\frac{1}{\beta}\ln(\Lambda_{0})}$
as a starting point. Therefore, we will proceed our discussion of
thermodynamic properties as a function of temperature and chemical
potential. Particularly, we will analyze entropy, internal energy,
correlation length, specific heat, electron density and isothermal
compressibility. We aim to study physical quantities around the pseudo-critical
temperature $T_{p}$. We stress that the present definition of
\textquotedbl pseudo-critical\textquotedbl{} is different to that
defined by Saito \citep{Saito} to describe the critical-like behavior in approaching the spinodal point near the first order transition. Therefore, by using a perturbation
approach, we can find the eigenvalues of transfer matrix \eqref{eq:V},
as discussed in Appendix \ref{Perturbative-Transfer-Matrix}.

In general, pseudo-transitions can be manipulated and analyzed using
perturbation techniques, as detailed in appendix     
\ref{Perturbative-Transfer-Matrix}.
For this purpose, we consider as unperturbed matrix $\mathbf{V}_{0}$
defined in appendix eq.\eqref{eq:transf-Mtx-0}. The eigenvalues of
the matrix \ref{eq:transf-Mtx-0}, are given by (\ref{eq:u0}-\ref{eq:u2}),
where we can observe the largest and second-largest eigenvalues are
given by \eqref{eq:u0} and \eqref{eq:u1}, respectively. However,
in low temperature region we have the condition $w_{0,2}\ll\{w_{0,0},w_{1,1},w_{2,2}\}$. This is so because the leading term of $w_{0,2}$ comes from the
lowest energy contribution of $w_{0,2}$ that includes only excited states, while the leading term of
$w_{0,0}$, $w_{1,1}$ and $w_{2,2}$ is given by the corresponding  ground state
energies of the system. Therefore, $w_{0,2}$ becomes exponentially small as compared to the other elements in the low-temperature regime. When \eqref{eq:u0} and \eqref{eq:u1}
become eventually equal, we can get a pseudo-critical temperature
from \eqref{eq:Psd-cond}, which reads 
\begin{equation}
u_{0}^{(0)}=u_{1}^{(0)}.\label{eq:trns-L0}
\end{equation}

Similar as discussed in \citep{pseudo}, the relation
\eqref{eq:trns-L0} can induce us to believe that there is a true
phase transition at finite temperature. However, the condition \eqref{eq:trns-L0}
does not mean that the transfer matrix eigenvalues satisfy $\Lambda_{0}=\Lambda_{1}$: the condition \eqref{eq:trns-L0}
is satisfied only when the matrix \eqref{eq:transf-Mtx-1} is ignored
and for $w_{0,2}=0$. Therefore, the first order perturbative corrections
(see eqs.\eqref{eq:u10-1} and \eqref{u11-1}) are $u_{0}^{(1)}>0$ and
$u_{1}^{(1)}<0$, which implies that the approximate solution given
in \eqref{eq:L_ap} must satisfy $\Lambda_{0}>\Lambda_{1}$.

\begin{figure}
\includegraphics[scale=0.5]{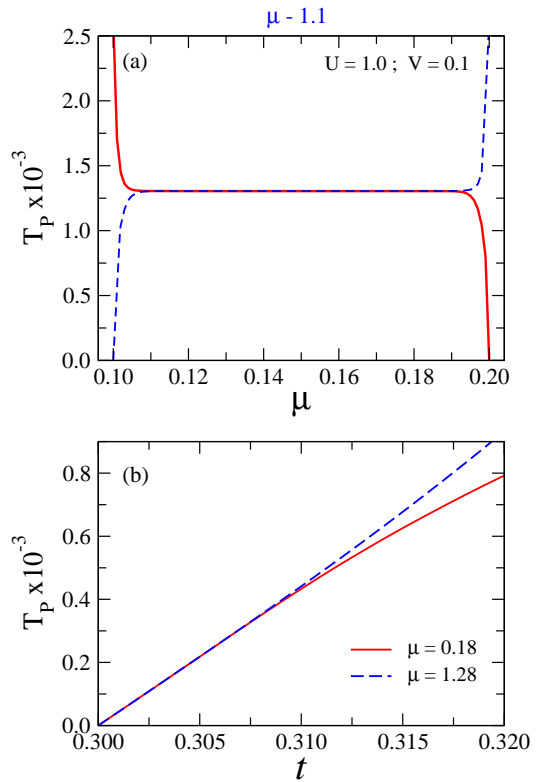}
\caption{\label{fig: crit-exp-T}(a) Pseudo-critical temperature $T_{p}\times10^{-3}$
for fixed parameters $t=0.303$, $U=1$ and $V=0.1$ near $AFM_{2}$
and $FRU_{2}$ phases boundary (solid line) and near the $AFM_{4}$
and $FRU_{4}$ phases boundary (dashed line). In the latter the chemical
potential was shifted to $\mu-1.1$ (top scale). (b) Pseudo-critical
temperature $T_{p}\times10^{-2}$ as a function of $t$ for $\mu=0.18$
(solid line) and $\mu=1.28$ (dashed line).}
\end{figure}

Returning to the relation \eqref{eq:trns-L0}, we get the following
equation 
\begin{equation}
w_{0,0}=2w_{1,1}\quad\text{for}\quad w_{0,0}>w_{2,2},
\end{equation}
and this one corresponds to the vicinity of the boundary line between $AFM_{2}$ and $FRU_{2}$. The pseudo-critical temperature $T_{p}$ is
illustrated in fig.\ref{fig: crit-exp-T}a where we plot  $T_{p}\times10^{-3}$ (red line) as a function of chemical
potential $\mu$, for fixed parameters $t=0.303$, $U=1$ and $V=0.1$.
In fig.\ref{fig: crit-exp-T}b we report $T_{p}$
for other values of $t$.

%\textcolor{red}{The explicit expression for the pseudo-critical condition of $t_{p}$ can be expressed from eq. \eqref{eq:w00} and \eqref{eq:wnn} we substitute $w_{0,0}=2w_{1,1}$, as follows}

%\textcolor{red}{
%\begin{equation}
%{\rm ch}(\tfrac{t_{p}}{T_{p}})=\frac{1+\frac{k_{p}x_{p}^{2}}{z_{p}^{2}}+\frac{x_{p}^{4}}{y_{p}^{4}z_{p}^{16}}-\frac{k_{p}x_{p}}{2}-\frac{1}{2x_{p}}-\frac{x_{p}^{3}}{2y_{p}^{4}z_{p}^{8}}}{2+\frac{2x_{p}^{2}}{y_{p}^{2}z_{p}^{4}}-\frac{4x_{p}}{z_{p}^{2}}\left(1+\frac{x_{p}^{2}}{y_{p}^{2}z_{p}^{8}}\right)},\label{eq:t_p}
%\end{equation}
%where 
%\begin{equation}
%k_{p}=\left(\varsigma_{p}+\varsigma_{p}^{-1}\right)y_{p}z_{p}+3y_{p}^{2}+z_{p}^{2},
%\end{equation}
%with $x_{p}$, $y_{p}$, $z_{p}$ and $\varsigma_{p}$ given below eq.\eqref{eq:w00} evaluated at $T_{p}$.}

Similarly, the relation \eqref{eq:trns-L0} leads to the following
equation 
\begin{equation}
w_{2,2}=2w_{1,1},\quad\text{for}\quad w_{0,0}<w_{2,2},\label{eq:w22,11}
\end{equation}
which holds in the close vicinity of the $FRU_4-AFM_4$ boundary-line. 

In fig.\ref{fig: crit-exp-T}a, we illustrate the pseudo-critical
temperature $T_{p}\times10^{-3}$ as a function of chemical potential
$\mu$. The continuous line corresponds to the pseudo-transition that
occurs at low chemical potentials for which there are nearly 2 electrons
per unit cell (near the boundary between $AFM_{2}$ and $FRU_{2}$
phases). The dashed line accounts for the second pseudo-transition
appearing at larger chemical potential for which there are 4 electrons
per unit cell in the ground-state (near the boundary between $AFM_{4}$
and $FRU_{4}$ phases). In this latter case, the chemical potential
was conveniently shifted to $\mu-V-U=\mu-1.1$ (top scale) to allow
representing both pseudo-transitions in the same frame.

In panel (b) the pseudo-critical temperatures $T_{p}\times10^{-2}$
as a function of $t$, for $\mu=0.18$ (red line) and $\mu=1.28$
(blue line) are illustrated.The pseudo-critical temperature vanishes linearly $T_p\propto (t-t_c)$ as one approaches the boundary-line $t_c=(U-V)/3$ from above, i.e., within one of the $AFM$ ground-states.

\subsection{Correlation length}

Since the eigenvalues are non-degenerate, the correlation length $\xi$
can be obtained by using the largest $\Lambda_{0}$ and second largest
$\Lambda_{1}$ eigenvalues given from \eqref{eq:sol-cub}. It is simply
written as

\begin{equation}\label{xi-def}
\xi=\left[\ln\left(\frac{\Lambda_{0}}{\Lambda_{1}}\right)\right]^{-1}.
\end{equation}

\begin{figure}
\includegraphics[scale=0.5]{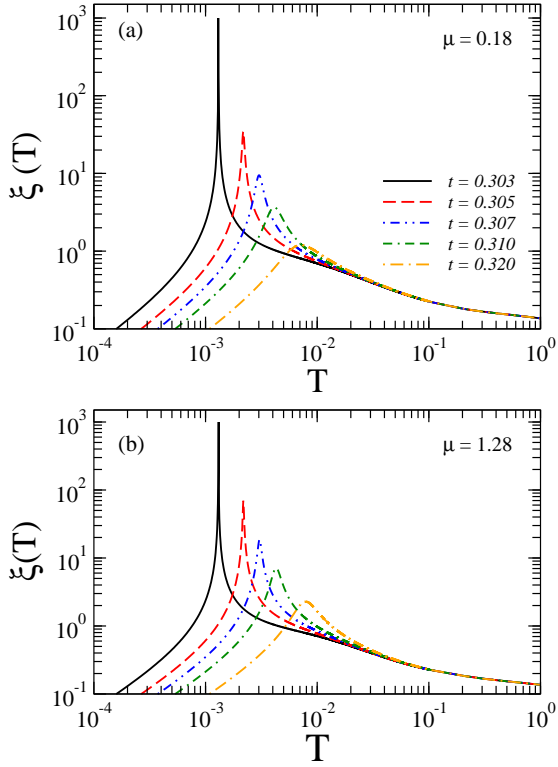}
\caption{\label{fig:Correlation}Correlation length as a function of temperature
in logarithmic scale for fixed $U=1$ and $V=0.1$.
(a) For several values of $t$, and fixed $\mu=0.18$ (in the vicinity of the $FRU_2-AFM_2$ phase-boundary). (b) For several
values of $t$, and fixed $\mu=1.28$ (in the vicinity of the $FRU_4-AFM_4$ phase-boundary). }
\end{figure}

In fig. \ref{fig:Correlation} we illustrate the correlation length
as a function of temperature. Panel (a) reports data for
several values of $t$ and assuming fixed $\mu=0.18$, $U=1$ and
$V=0.1$. Here we observe how the peak becomes more pronounced when $t\rightarrow t_c=(U-V)/3=0.3$ (approaching the $FRU_2-AFM_2$ ground-state phase-boundary). For
larger $t$ the height of the peak becomes lower and broader. For $t=0.303$ we already observe
a strong peak around $T_{p}$, which was computed with high precision
to be $T_{p}=1.3046184136496\times10^{-3}$. Similarly, panel (b)
reports for the same parameters set used in (a) but for $\mu=1.28$ (in the vicinity of the $FRU_4-AFM_4$ phase-boundary).

There are two situations where pseudo-transition occurs. The first
one satisfy the condition $w_{0,0}\sim2w_{1,1}$ but in the perturbation regime for which $8w_{0,1}^2\ll(w_{0,0}-2w_{1,1})^2$. Under this condition, we have the following result,

\begin{equation}\label{L0/L1}
\frac{\Lambda_{0}}{\Lambda_{1}}\rightarrow\begin{cases}
\frac{w_{0,0}}{2w_{1,1}}, & w_{0,0}>2w_{1,1}\\
\frac{2w_{1,1}}{w_{0,0}}, & w_{0,0}<2w_{1,1}
\end{cases}.
\end{equation}

Consequently, in the regime where the perturbation approach stands,  the correlation length in the close vicinity of the pseudo-critical temperature
can be expressed by

\begin{equation}\label{xi-limit_0}
\xi_{0}(\tau)=c_{0,\xi}\left|\tau\right|^{-1}+{\cal O}(\tau^{0}) .
\end{equation}
Here $\tau=(T-T_{p})/T_{p}$ and the coefficient is given by

\begin{equation}
c_{0,\xi}=\frac{\tilde{w}_{0,0}}{T_{p}\tilde{v}_{0}},
\end{equation}
with $\tilde{w}_{0,0}$ is $w_{0,0}$ evaluated at $T=T_{p}$, and
\begin{equation}
\tilde{v}_{0}=\left|\frac{\partial\left(w_{0,0}-2w_{1,1}\right)}{\partial T}\right|_{T_{p}}.
\end{equation}

Similarly, the second pseudo-transition occurs when $w_{2,2}\sim2w_{1,1}$
but in the perturbation regime $8w_{1,2}^2\ll(w_{2,2}-2w_{1,1})^2$, that
provides us 
\begin{equation}
\frac{\Lambda_{0}}{\Lambda_{1}}\rightarrow\begin{cases}
\frac{w_{2,2}}{2w_{1,1}}, & w_{2,2}>2w_{1,1}\\
\frac{2w_{1,1}}{w_{2,2}}, & w_{2,2}<2w_{1,1}
\end{cases}.
\end{equation}

Analogously, the correlation length close to the pseudo-transition, can
be expressed around $T_{p}$, resulting in

\begin{equation}\label{xi-limit_2}
\xi_{2}(\tau)=c_{2,\xi}\left|\tau\right|^{-1}+{\cal O}(\tau^{0}),
\end{equation}
where the coefficient is given by

\begin{equation}
c_{2,\xi}=\frac{\tilde{w}_{2,2}}{T_{p}\tilde{v}_{2}}.
\end{equation}
with $\tilde{w}_{2,2}$ evaluated at $T=T_{p}$, and 
\begin{equation}
\tilde{v}_{2}=\left|\frac{\partial\left(w_{2,2}-2w_{1,1}\right)}{\partial T}\right|_{T_{p}}.
\end{equation}
These pre-asymptotic power-laws fail for very small
$\tau$, because the perturbation conditions can not be satisfied at $T_p$. Therefore, the correlation length actually depicts a pronounced peak and not a true divergence. 

%In both limits, we are considering $w_{0,2}<\{w_{0,0},w_{1,1},w_{2,2}\}$.

\begin{figure}
\includegraphics[scale=0.5]{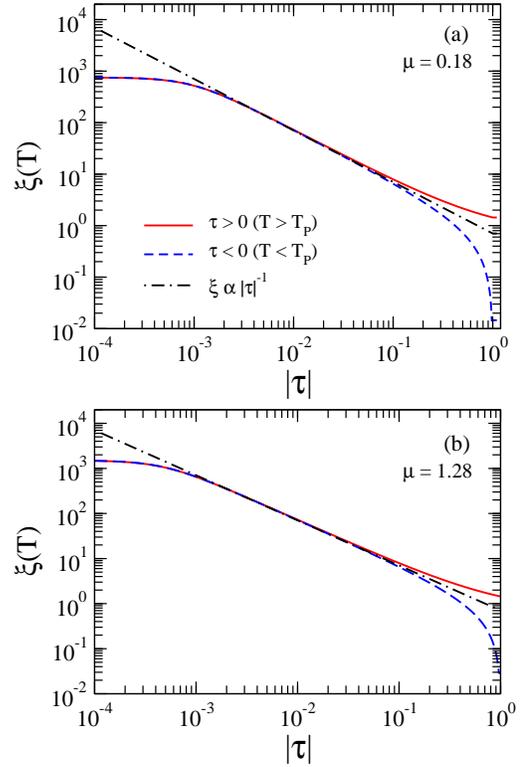}

\caption{\label{fig:Corre-length-exp}Correlation length as a function of $\tau$,
for fixed parameters $U=1$, $V=0.1$, $t=0.303$. Red line corresponds
for $\tau>0$ ($T>T_{p}$) and dashed blue line denotes $\tau<0$
($T<T_{p}$), while straight dash-doted line reports $\xi\propto|\tau|^{-1}$.
(a) For $\mu=0.18$. (b) For $\mu=1.28$. }
\end{figure}

In fig.\ref{fig:Corre-length-exp}a the correlation length is depicted
as a function of $\tau$ in logarithmic scale, assuming fixed parameters
$U=1$, $V=0.1$, $t=0.303$ and $\mu=0.18$. The solid curve represents
$\tau>0$ ($T>T_{p}$) while the dashed curve denotes $\tau<0$ ($T<T_{p}$).
The straight dash-doted line reports the asymptotic limit $\xi\propto|\tau|^{-1}$,
where we observe clearly the critical exponent $\nu=1$ in the pseudo-critical
regime. Notice that when $T$ is very close to $T_{p}$ the power
law critical exponent fails, evidencing the ultimate non-singular
behavior of the thermodynamic quantities. A similar plot is depicted
in fig. \ref{fig:Corre-length-exp}b, assuming the same set of parameters
but for chemical potential $\mu=1.28$. Here again we observe manifestly
the critical exponents $\nu=1$. Note that the power-law behavior holds for nearly two decades.

In order to have a clear picture of the range of temperatures around $T_p$ on which the pseudo-critical regime holds, we can estimate crossover boundaries between the non-singular and power-law regimes, as well as between the power-law regimes and the low and high temperature ones. The non-singular regime in the close vicinity of $T_p$ emerges as the perturbation analysis fails.  In the vicinity of the $FRU_2-AFM_2$ ground-state transition,  crossover lines can be built using $8w_{0,1}^2=(w_{0,0}-2w_{1,1})^2$, which separates the perturbative from the non-perturbative regimes. A similar crossover line can be written in the vicinity of the $FRU_4-AFM_4$ ground-state transition [$8w_{1,2}^2=(w_{2,2}-2w_{1,1})^2$]. The power-law regime holds while the correlation length remains much larger the  lattice spacement. We will consider that crossover lines from the power-law to the low and high-temperature regimes satisfies $\xi=10$. Eqs. \eqref{xi-def}  and \eqref{L0/L1}  can be used to draw the respective crossover lines.

\begin{figure}
\includegraphics[scale=0.60]{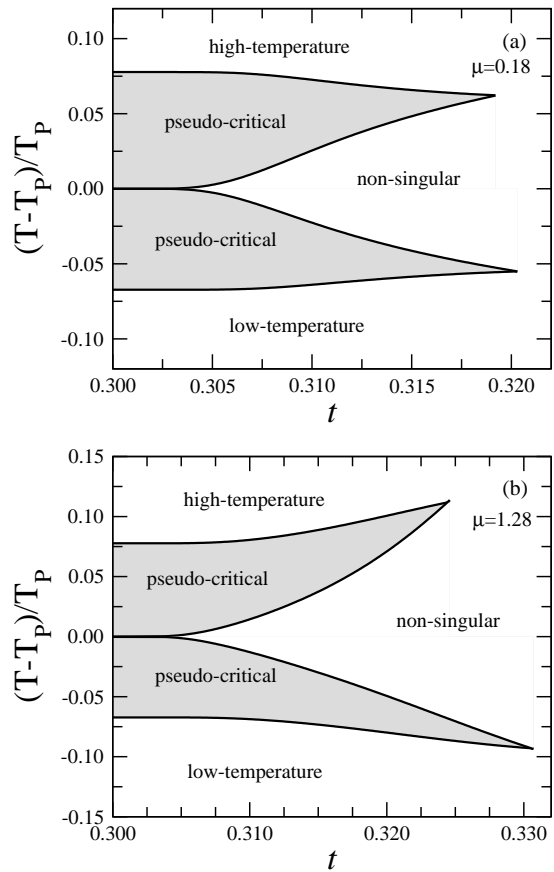}

\caption{\label{crossover} Distinct temperature regimes as a function of the hopping parameter $t$ in the close vicinity of the $FRU_2-AFM_2$(a) and $FRU_4-AFM_4$(b) ground-state transitions. Here we used $U=1$, $V=0.3$, (a) $\mu=0.18$ and (b) $\mu=1.28$. The non-singular regime becomes exponentially narrow as $t\rightarrow t_c=(U-V)/3=0.3$. The power-law regime holds for $\tau=(T-T_p)/T_p$ of the order of a few percents in the close vicinity of $t_c$. It dies away far from $t_c$ when the perturbation and large correlation length conditions become incompatible.}
\end{figure}

In fig.\ref{crossover} we show the above crossover lines as a function the hopping parameter $t$ in the close vicinity's of the $FRU_2-AFM_2$ and $FRU_4-AFM_4$ boundary lines for an illustrative set of the other Hamiltonian parameters. Firstly notice that the width of the non-singular regime decreases quite fast as one approaches $t_c=(U-V)/3$.  While the pseudo-transition temperature $T_p$ decreases linearly, the non-singular temperature range decays much faster as $e^{-a/(t-t_c)}$, with $a$ being a constant at $t_c$. The non-singular regime widens as we further depart from $t_c$. On the other hand, the crossover lines to the high and low-temperature regimes are weakly dependent on $t$. Therefore, in the close vicinity of $t_c$, the power-law regime extends over a temperature range of the order of a few percents of $T_p$. The crossover lines delimiting the non-singular and the high and low temperature regimes eventually meet. After this point, no pre-asymptotic power-law regime can be identified.

\section{First-order like pseudo-transition }

It is interesting to analyze some physical quantities, which are obtained
as a first derivative of the free energy and exhibit an almost step-like
behavior.

\subsection{Entropy }

\begin{figure}
\includegraphics[scale=0.95]{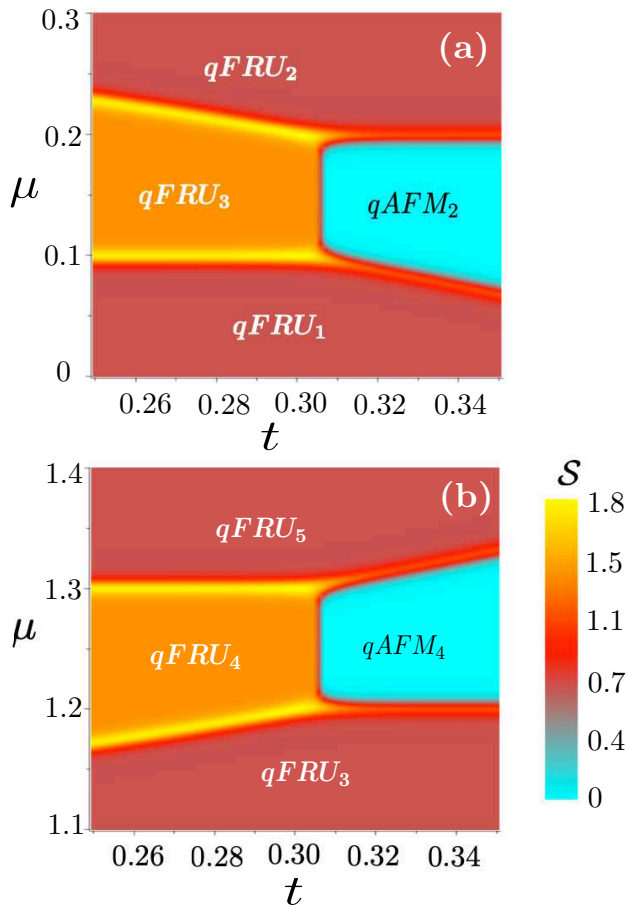}
\caption{\label{fig:Density-S}Density plot of entropy in the plane $t-\mu$,
assuming fixed $V=0.1$, $U=1$ and $T=0.005$. (a) For $\mu=[0,0.3]$.
(b) For $\mu=[1.1,1.4]$.}
\end{figure}

In fig.\ref{fig:Density-S}a we illustrate the density plot of entropy
(${\displaystyle \mathcal{S}=-\frac{\partial f}{\partial T}}$ ) in
the plane $t-\mu$, for fixed parameters $V=0.1$, $U=1$ and $T=0.005$,
for $\mu=[0,0.3]$. This plot is depicted in the same scale of
the phase diagram illustrated in fig.\ref{fig:Phase-diagram}a. We
definitely observe a pseudo-transition in the boundary of quasi-antiferromagnetic
$qAFM_{2}$ and quasi-frustrated $qFRU_{2}$ phases characterized
by a sharp boundary, while electron density remains constant $\rho=2/3$.
The regions are mostly governed by the zero temperature phases but,
due to thermal fluctuations, the zero temperature phases become \textquotedbl quasi\textquotedbl{}
long-ranged because of the lack of actual spontaneous long-range order
at any finite temperature. A similar density plot is illustrated in
fig.\ref{fig:Density-S}b, assuming the same parameters but in the
interval of $\mu=[1.1,1.4]$. We can observe a steepy entropy
change in the boundary between the $qAFM_{4}$ and $qFRU_{4}$ phases
where the electron density $\rho=4/3$ per cell remains unaltered.

\begin{figure}
\includegraphics[scale=0.50]{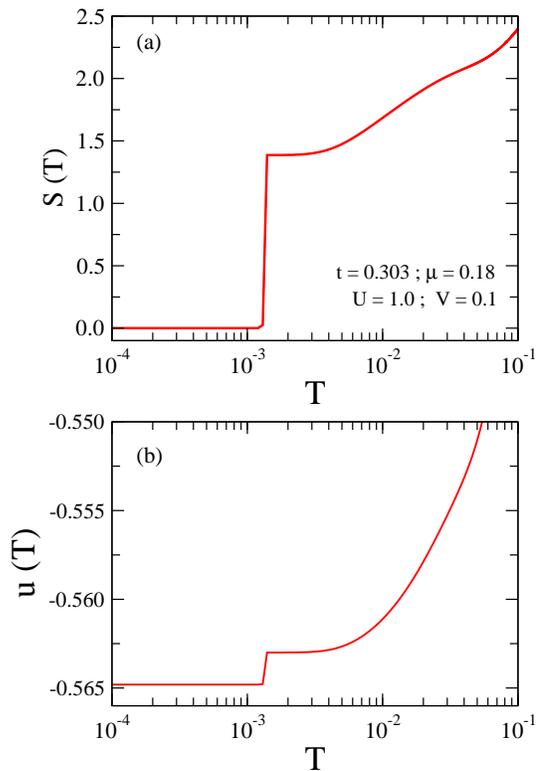}
\caption{\label{fig:Entropy-energ}(a) Entropy as a function of temperature
in semi-logarithmic scale, for fixed $t=0.303$, $\mu=0.18$, $U=1$
and $V=0.1$. (b) Internal energy as a function of temperature in
semi-logarithmic scale, for fixed $t=0.303$, $\mu=0.12$, $U=1$ and
$V=0.1$.}
\end{figure}

In fig.\ref{fig:Entropy-energ}a we report the magnitude of entropy
as a function of temperature in semi-logarithmic scale, assuming fixed
parameters $t=0.303$, $\mu=0.18$, $U=0.1$. Clearly one can observe
a nearly step behavior at $T_{p}=1.3\times10^{-3}$. For temperatures
below $T_{p}$ the entropy is almost null, which corresponds to the
quasi-antiferromagnetic phase ($qFAM_{2}$). On the other hand, the
entropy has a plateau region ($qFRU_{2}$) with $\mathcal{S}_{0}=\ln(4)$
for $T>T_{p}$. As the temperature is further increased, the entropy
behaves like in standard models. Therefore we observe a strong continuous
change of entropy around the pseudo-transition, which typically resembles
a discontinuous (first-order) phase transition. However, we stress
that it remains analytical at $T_{p}$. 
%\textcolor{red}{At first glance, one can try to define a latent heat as tried to define in ref. \citep{pseudo}, but as there is no real jump, this quantity cannot be defined. It is worth noting that it is not possible to define the validity of pseudo-transition as a function of temperature\citep{pseudo,hutak21}. Since the height of the pseudo-transition peak decreases smoothly with pseudo-critical temperature.}

\subsection{Internal energy}

Other quantity we discuss here is the internal energy $u(T)=T{\cal S}+f$,
which can also be obtained after a first-derivative of the free-energy.

In fig.\ref{fig:Entropy-energ}b, the internal energy as a function
of temperature in semi-logarithmic scale is depicted, for fixed $t=0.303$,
$\mu=0.18$, $U=1$ and $V=0.1$. Internal energy is continuous as
a function of temperature, although it also shows a steep variation
around the pseudo-critical temperature $T_{p}$. For temperatures
below $T_{p}$, the internal energy leads to $u=-0.5648$, which corresponds
to the $AFM_{2}$ ground state energy $E_{AFM_2}$. For $T>T_{p}$, the
internal energy is mostly dominated by the $FRU_{2}$ ground-state energy $E_{FRU_2}=-0.563$.

\subsection{The electron density}

Another important quantity we explore is the thermal average electron
density ${\displaystyle \rho=-\Bigl(\frac{\partial f}{\partial\mu}\Bigr)}$
per unit cell.

\begin{figure}
 \resizebox{8.5cm}{6.0cm}{\includegraphics{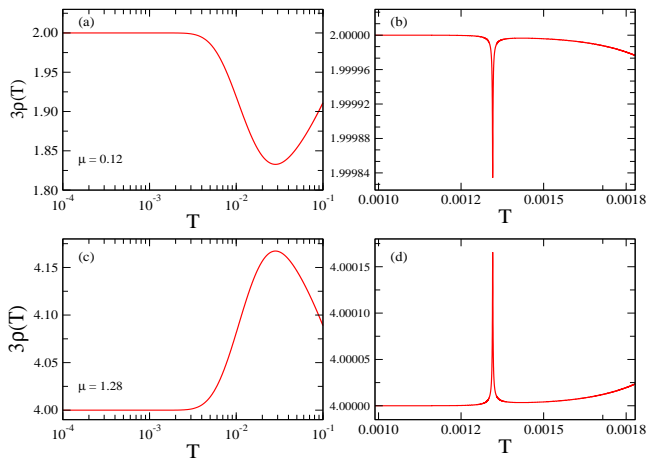}}

\caption{\label{fig:E-d mu fixed t}Electron density as a function of temperature
in semi-logarithmic scale, for fixed $t=0.303$, $U=1$ and $V=0.1$.
(a) For $\mu=0.12$. (b) Zooming panel (a) around $T_{p}$. (c) For
$\mu=1.28$. (d) Zooming panel (c) around $T_{p}$. }
\end{figure}

In fig.\ref{fig:E-d mu fixed t} the electron density as function
of temperature is reported in semi-logarithmic scale, for fixed $t=0.303$,
$U=1$, $V=0.1$ and $\mu=0.12$. At low temperatures, the electron
density remains almost constant $\rho=2/3$, up to roughly around
$T\sim0.01$, and then the electron density decreases to $3\rho\approx1.80$,
for higher temperature the electron density increases with temperature
( for further details see reference \citep{hubd-dmd-ch}). Therefore
apparently no pseudo-critical behavior at $T_{p}$ is evidenced in
$\rho$. In fig.\ref{fig:E-d mu fixed t}b we plot the electron density
as a function of temperature around pseudo-critical temperature. We
observe a tiny depression at $T_{p}$ reminiscent of the pseudo-transition due to thermal excitations to states with a single electron per unit cell (see phase-diagram). 

Similarly, the electron density is depicted in fig.\ref{fig:E-d mu fixed t}c
as function of temperature in semi-logarithmic scale, assuming same
set of parameters, but for $\mu=1.28$. Once again electron density
remains almost constant $\rho=4/3$, for temperatures below $T\sim0.01$,
while for higher temperature there is a peak reaching $\rho\approx4.20$.
In fig.\ref{fig:E-d mu fixed t}d we present a zooming plot of panel
(c) around the pseudo-critical temperature, where a tiny peak reminiscent
of the pseudo-critical temperature is also present refleting thermal excitations to states with 5 electrons per unit cell (see phase diagram). Surely we cannot
expect any strong change in the electron density around $T_{p}$ because
the two competing grounds state have equivalent electron densities.

%It is worth mentioning that the first derivative of free energy leads us to believe that we are facing a first-order phase transition.

\section{Second-order like pseudo-transition}

Now let us turn our attention to the second-order derivative of the
free energy. The following quantities exhibit trends quite similar
to second-order phase transition. It is important to reinforce that
there is no singularity at $T_{p}$ but just a rather sharp peak.

\subsection{The specific heat}

\begin{figure}
\includegraphics[scale=0.5]{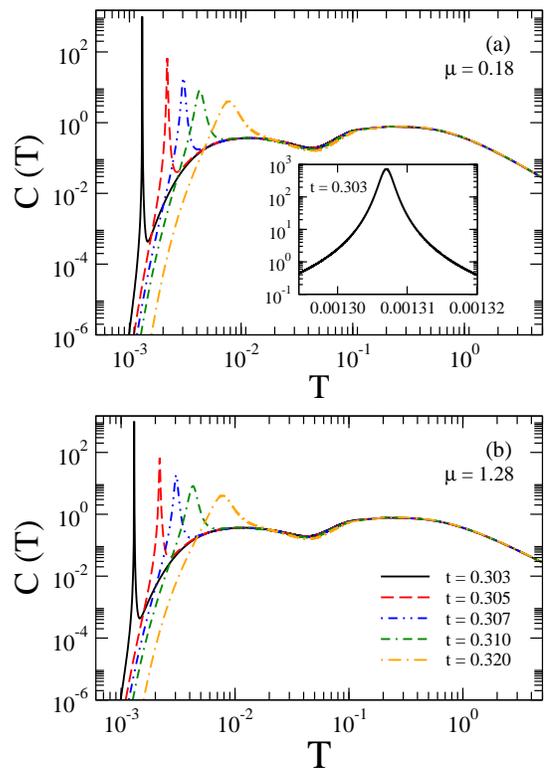}

\caption{\label{fig:Specific-heat-as}Specific heat as a function of temperature
in logarithmic scale for $U=1$ and $V=0.1$. (a)
For several values of $t$ and $\mu=0.18$ (in the vicinity of the $FRU_2-AFM_2$ phase-boundary). Inset: zooming around $T_{p}$. (b) For several
values of $t$ and $\mu=1.28$ (in the vicinity of the $FRU_4-AFM_4$ phase-boundary).} 
\end{figure}

In fig.~\ref{fig:Specific-heat-as}, we display the specific heat
${\displaystyle C=T\Bigl(\frac{\partial\mathcal{S}}{\partial T}\Bigr)}$
as a function of temperature in logarithmic scale assuming
fixed $U=1$, $V=0.1$. In panel (a) we depict for
several values of $t$ and $\mu=0.18$ (close to the $FRU_2-AFM_2$ ground-state transition). Here we see how the height of the peak increases and the peak becomes sharper when $t\rightarrow0.3$, while for $t$
larger the height of the peak becomes lower and broader. At low
temperatures, we observe clearly a huge sharp peak quite similar to
a second-order phase transition divergence around $T_{p}$. However,
a zooming look as provided in the inner plot around
$T_{p},$ evidences the rounded nature of the peak at the
pseudo-critical temperature. Panel (b)  reports the specific heat
for the same set of parameters used in panel (a) but for $\mu=1.28$ (close to the $FRU_4-AFM_4$ ground-state phase transition).

Now let us analyze the nature of the peak around $T_{p}$, looking
for some critical exponent universality. For this purpose, we consider
the specific heat around pseudo-critical temperature in asymptotic
limit (but not very close to $T_{p}$), which can be expressed according
the discussion in reference \citep{unv-cr-exp} as

\begin{equation}
C(\tau)=T\left.\left(\frac{\partial\mathcal{S}}{\partial\tau}\right)\left(\frac{\partial\tau}{\partial T}\right)\right|_{T_{p}}=2c_{f}\tau^{-3},
\end{equation}
where the coefficient $c_{f}$ is given by 
\begin{equation}
c_{f}=\begin{cases}
\tilde{w}_{0,1}^{2}c_{0,\xi}, & w_{0,0}\sim2w_{1,1}\\
\tilde{w}_{1,2}^{2}c_{2,\xi}, & w_{2,2}\sim2w_{1,1}
\end{cases}.
\end{equation}

Therefore the specific heat has the following pre-asymptotic expression when approaching the pseudo-critical temperature

\begin{equation}
C(\tau)\propto|\tau|^{-\alpha},
\end{equation}
with critical exponent $\alpha=3$. The Hubbard diamond chain in the
atomic limit also satisfy the pseudo-critical exponent found in reference
\citep{unv-cr-exp}. However, it is worth stressing
that this pre-asymptotic regime is only valid around the ascending and
descending parts of the peak, and surely fails very close to the peak top where the perturbation condition can not prevail.

\begin{figure}
\includegraphics[scale=0.5]{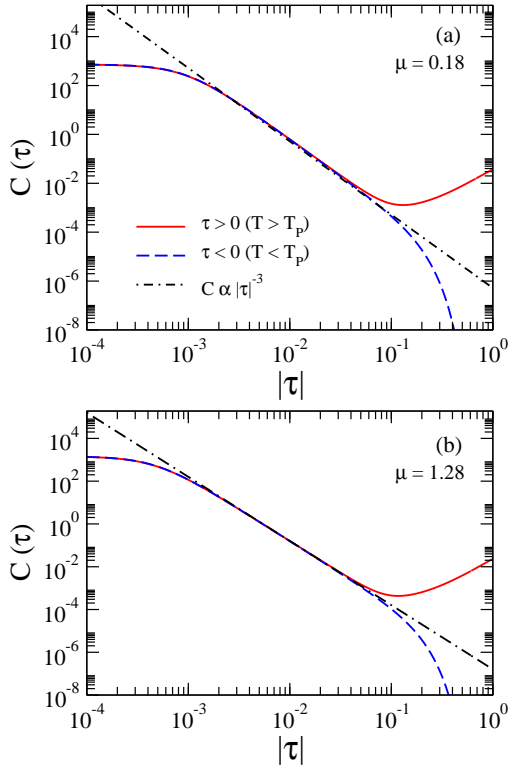}\caption{\label{fig:crit-exp}Logarithmic specific heat as a function of $\ln(|\tau|)$,
for fixed parameters $U=1$, $V=0.1$, $t=0.303$. Solid line corresponds
for $\tau>0$ ($T>T_{p}$) and dashed line denotes $\tau<0$ ($T<T_{p}$).
The straight dash-doted line reports $\xi\propto|\tau|^{-3}$. (a)
For $\mu=0.18$. (b) For $\mu=1.28$. }
\end{figure}

In order to confirm the above result. We report in fig.\ref{fig:crit-exp}a
the $C(\tau)$ as a function of $\tau$, assuming fixed
parameters $U=1$, $V=0.1$, $t=0.303$ and $\mu=0.18$ (near the
$AFM_{2}-FRU_{2}$ phase-boundary). Continuous line represents data
above $T_{p}$, while dashed line corresponds data below $T_{p}$.
The dash-doted line describes the asymptotic function, with critical
exponent $\alpha=3$. The straight line with angular coefficient
$\alpha=3$ fits accurately data over nearly two decades. Very close
to $T_{p}$ the power-law behavior breaks down, reflecting the actual
analytic behavior of the specific heat. In panel (b) we plot  the specific heat for $\mu=1.28$ (near the $AFM_{4}-FRU_{2}$
phase boundary), showing similar trends.

\subsection{Compressibility}

Another interesting quantity to be discuss is the isothermal electron
compressibility\citep{baxter-book} ${\displaystyle \kappa_{T}=\frac{1}{\rho^{2}}\Bigl(\frac{\partial\rho}{\partial\mu}\Bigr)_{T}}$,
as a function of Hamiltonian parameters, temperature and electron
density.

\begin{figure}
\includegraphics[scale=0.5]{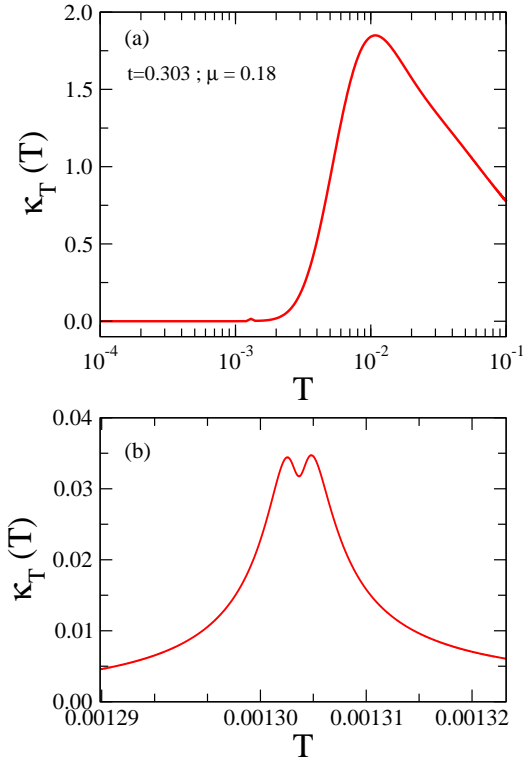}\caption{\label{fig:Isotermal-compressibility-as}Isothermal compressibility
as a function of temperature in semi-logarithmic scale, for fixed
$t=0.303$, $\mu=0.18$, $U=1$ and $V=0.1$. (b) Magnified around
$T_{p}$.}
\end{figure}

In fig.\ref{fig:Isotermal-compressibility-as}a the isothermal compressibility
is shown as a function of temperature in semi-logarithmic scale, we
observe a tiny peak at the pseudo-critical temperature. Magnifying
around the pseudo-critical temperature (see fig. \ref{fig:Isotermal-compressibility-as}b)
it illustrated a double peak with local minimum at $T_{p}$. Since
the pseudo-transition results from competing ground-states with the
same electron density, we actually cannot expect any giant peak of
the electron compressibility at $T_{p}$ because that would result in
a pronounced electron density change.

\section{Summary and Conclusions}

In summary, we considered the extended Hubbard diamond chain restricted
to the atomic limit with an arbitrary number of particles driven by
chemical potential. The interaction between dimer diamond chain and
nodal couplings are taken in the atomic limit (no hopping), while
dimer interaction includes the hopping term. We showed that this model
exhibits a pseudo-transition effect in the low-temperature region.
The internal energy and entropy were shown to change quite abruptly
in a very narrow range of temperatures around the pseudo-transition
when the physical parameters are properly tuned in a close vicinity
of special ground-state phase transitions. The correlation length
and specific heat display pronounced peaks with well defined power-laws
in a well defined temperature range in the vicinity of the pseudo-transition
point. The pseudo-critical exponents associated with the correlation
length and specific heat were shown to be $\nu=1$ and $\alpha=3$,
respectively. These are the same pseudo-exponents reported to hold
is spin models with Ising-like interactions, pointing towards a universal
behavior of pseudo-transitions\cite{unv-cr-exp,hutak21}. We also demonstrated that the electron
density and respective electronic compressibility displays reminiscent
signatures of the pseudo-transition. The present results add to the
general understanding of the remarkable phenomenon of pseudo-transitions
taking place at finite temperatures in one-dimensional equilibrium
systems having structured interactions within the relevant unit cell.

\section{Acknowledgements}

This work was supported by CAPES (Coordenação de Aperfeiçoamento de
Pessoal de Nível Superior), CNPq (Conselho Nacional de Desenvolvimento
Científico e Tecnológico), FAPEAL (Fundação de Apoio à Pesquisa do
Estado de Alagoas), and FAPEMIG (Fundação de Apoio à Pesquisa do Estado
de Minas Gerais).

\appendix
%dummy comment inserted by tex2lyx to ensure that this paragraph is not empty

\section{Real roots of cubic equation\label{Real-roots-of}}

Transfer matrix is a symmetric matrix, so its eigenvalues are guaranteed
to be real values. Therefore, here we verify the cubic equation \eqref{eq:cub-eq}
roots must be different and real numbers, given by \eqref{eq:sol-cub}.

Obviously, we can convince that both $Q$ and $R$ are real numbers
from \eqref{eq:Q real} and \eqref{eq: R real}.

Now let verify the real number $Q$ is positively defined by using
the cubic equation coefficients \eqref{eq:coefs}, thus after some
algebraic manipulation we obtain the following expression 
\begin{alignat}{1}
Q= & \tfrac{1}{9}\left(w_{2,2}-\tfrac{1}{2}w_{0,0}-w_{1,1}\right)^{2}+\tfrac{1}{12}\left(w_{0,0}-2w_{1,1}\right)^{2}\nonumber \\
 & +\tfrac{2}{3}w_{0,1}^{2}+\tfrac{1}{3}w_{0,2}^{2}+\tfrac{2}{3}w_{1,2}^{2},\label{eq:Q}
\end{alignat}
which is strictly a positive number ($Q>0$), since all Boltzmann
factors are positive. According to a cubic equation property, this
condition is enough to conclude that three the eigenvalues of transfer
matrix will be different and real numbers.

Now the question is to determine which are the largest and the lowest
eigenvalues? Since all eigenvalues must be real and different because
$Q>0$. Which implies that the solution \eqref{eq:sol-cub}, must
satisfy the following restriction $\phi\ne\pm n\pi$, with $n=\{0,1,2,\dots\}$.
Therefore, we can choose $\phi$ conveniently such that $0<\phi<\pi$.
Thus we have the following trigonometric relation relation

\begin{equation}
\cos\left(\tfrac{\phi}{3}\right)>\cos\left(\tfrac{\phi-2\pi}{3}\right)>\cos\left(\tfrac{\phi-4\pi}{3}\right),
\end{equation}
if we multiply all terms of inequalities by $2\sqrt{Q}>0$, and adding
$-\frac{a_{3}}{3}$ to all expressions (note that $a_{3}<0$), thus
we conclude that

\begin{equation}
\Lambda_{0}>\Lambda_{1}>\Lambda_{2}.
\end{equation}
Note that if $\phi=0$, in principle we would have $\Lambda_{1}=\Lambda_{2}$
but this condition is forbidden because $Q>0$. Similarly for $\phi=\pi$
we have $\Lambda_{0}=\Lambda_{1}$, but again this condition cannot
be satisfied since $Q>0$. Therefore, we have identified which eigenvalues
is the largest one and the lowest one.

Of course we can choose other intervals equivalently, and verify how
the eigenvalues are ordered. In table \ref{tab:Cubic-r} is reported
the eigenvalues for each intervals. The open intervals of $\phi$
guaranties all eigenvalues must be real and different values, in order
to satisfy the condition $Q>0$.

\begin{table}
\begin{tabular}{|c|c|c|c|}
\hline 
$\phi$  & Largest  & 2nd Largest  & Lowest\tabularnewline
\hline 
\hline 
$\langle0,\pi\rangle$  & $\Lambda_{0}$  & $\Lambda_{1}$  & $\Lambda_{2}$\tabularnewline
\hline 
$\langle\pi,2\pi\rangle$  & $\Lambda_{1}$  & $\Lambda_{0}$  & $\Lambda_{2}$\tabularnewline
\hline 
$\langle2\pi,3\pi\rangle$  & $\Lambda_{1}$  & $\Lambda_{2}$  & $\Lambda_{0}$\tabularnewline
\hline 
$\langle3\pi,4\pi\rangle$  & $\Lambda_{2}$  & $\Lambda_{1}$  & $\Lambda_{0}$\tabularnewline
\hline 
$\langle4\pi,5\pi\rangle$  & $\Lambda_{2}$  & $\Lambda_{0}$  & $\Lambda_{1}$\tabularnewline
\hline 
$\langle5\pi,6\pi\rangle$  & $\Lambda_{0}$  & $\Lambda_{2}$  & $\Lambda_{1}$\tabularnewline
\hline 
\end{tabular}\caption{\label{tab:Cubic-r}Cubic root solutions are tabulated in decreasing
order by intervals in $\phi$.}
\end{table}

The three cubic root solutions are exchanging periodically which depends
on interval of $\phi$, making a bit puzzle to identify which eigenvalues
is the largest one. In table \ref{tab:Cubic-r} it is reported how
the eigenvalues are ordered. In each interval the solutions are equivalent,
because other intervals simply exchange the eigenvalues with no relevance.
Therefore, here we conveniently choose the first interval $\phi\in\langle0,\pi\rangle$,
so the eigenvalues must be ordered as follow $\Lambda_{0}>\Lambda_{1}>\Lambda_{2}$.

Even more, according to Perron-Frobenius theorem, the largest eigenvalue
must be non-degenerate and positive.

\section{Perturbative Transfer Matrix Correction\label{Perturbative-Transfer-Matrix}}

In order to study the pseudo-transitions property in low-temperature
region, we need to analyze the transfer matrix in the low temperature
region. In principle, cubic root solution could be a bit cumbersome
task to identify as an easy handling solutions. A simple strategy
to obtain a reasonable solution could be considering the transfer
matrix \eqref{eq:transf-Mtx}, as a sum o two matrix $\mathbf{V}=\mathbf{V}_{0}+\zeta\,\mathbf{V_{1}}$.
The first term is given by 
\begin{equation}
{\bf V}_{0}=\left[\begin{array}{ccc}
w_{0,0} & 0 & w_{0,2}\\
0 & 2w_{1,1} & 0\\
w_{0,2} & 0 & w_{2,2}
\end{array}\right],\label{eq:transf-Mtx-0}
\end{equation}
This matrix can be considered as the unperturbed transfer matrix,
whose eigenvalues are

\begin{alignat}{1}
u_{0}^{(0)}= & \tfrac{1}{2}\left[w_{0,0}+w_{2,2}+s\right],\label{eq:u0}\\
u_{1}^{(0)}= & 2w_{1,1},\label{eq:u1}\\
u_{2}^{(0)}= & \tfrac{1}{2}\left[w_{0,0}+w_{2,2}-s\right],\label{eq:u2}
\end{alignat}
where $s=\sqrt{d^{2}+4w_{0,2}^{2}}$ and $d=w_{0,0}-w_{2,2}$.

In the limit of $w_{0,2}\rightarrow0$, the unperturbed solution can
eventually satisfy the following relation 
\begin{equation}
u_{0}^{(0)}=u_{1}^{(0)}.\label{eq:Psd-cond}
\end{equation}
This condition leads to a pseudo-transition, which we can simplify:
\begin{alignat}{1}
w_{0,0}=2w_{1,1},\quad\text{for}\quad w_{0,0}>w_{2,2},\\
w_{2,2}=2w_{1,1},\quad\text{for}\quad w_{0,0}<w_{2,2}.
\end{alignat}
The above condition is essential to find pseudo-critical temperature
$T_{p}$.

The second term of transfer matrix is given by

\begin{equation}
{\bf V}_{1}=\left[\begin{array}{ccc}
0 & \sqrt{2}\,w_{0,1} & 0\\
\sqrt{2}\,w_{0,1} & 0 & \sqrt{2}\,w_{1,2}\\
0 & \sqrt{2}\,w_{1,2} & 0
\end{array}\right].\label{eq:transf-Mtx-1}
\end{equation}

Close to the pseudo-critical temperature, the elements of matrix $\mathbf{V}_{1}$
become smaller than the elements of matrix $\mathbf{V}_{0}$. Therefore,
we can find the transfer matrix eigenvalues by using a perturbative
approach.

Consequently, the solution of the transfer matrix after a standard
perturbative manipulation up to first order term gives us the following
root corrections,

\begin{alignat}{1}
u_{0}^{(1)}= & \frac{\left[(s-d)w_{1,2}+2w_{0,2}w_{0,1}\right]^{2}}{s(s-d)\left(u_{0}^{(0)}-2w_{1,1}\right)},\label{eq:u10-1}\\
u_{1}^{(1)}= & -u_{0}^{(1)}-u_{2}^{(1)},\label{eq:u12-1}\\
u_{2}^{(1)}= & \frac{\left[(s+d)w_{1,2}-2w_{0,2}w_{0,1}\right]^{2}}{s(s+d)\left(u_{2}^{(0)}-2w_{1,1}\right)}.\label{u11-1}
\end{alignat}

As a consequence of the perturbative correction, the transfer matrix
eigenvalues becomes

\begin{alignat}{1}
\Lambda_{j}= & u_{j}^{(0)}+\zeta\,u_{j}^{(1)}+\mathcal{O}(\zeta^{2}),\quad j=\{0,1,2\}.\label{eq:L_ap}
\end{alignat}

Assuming the elements of matrix $V_{1}$ are small, we can obtain
an approximate result of~(\ref{eq:sol-cub}), fixing $\zeta=1$.
Hence, it is evident that the largest eigenvalue of the transfer matrix
will be $\Lambda_{0}$, even when $u_{0}^{(0)}=u_{1}^{(0)},$ since
$u_{0}^{(1)}>0$ and $u_{1}^{(1)}<0$.

\end{document}